\documentclass[a4paper,11pt]{article}
\usepackage{pos}

\title{Implementation of the three-neutron quantization condition}
\author*[a]{Wilder Schaaf}
\author[a]{Stephen R. Sharpe}

\affiliation[a]{Dept. of Physics, University of Washington,\\
Seattle, WA, USA}

\emailAdd{wschaaf@uw.edu}
\emailAdd{srsharpe@uw.edu}

\abstract{
We present an implementation of the three-neutron quantization condition (QC) derived in previous work. We construct the matrices appearing in the QC and determine solutions numerically. The symmetries of the QC allow the projection onto irreducible representations of the appropriate little group (depending on frame momentum), restricting the size of the matrices and reducing computational complexity. 
In this initial study, we include only two-neutron interactions, which are modeled based on experimental data for $I=1$ scattering amplitudes. 
We show examples of the finite-volume spectrum in two frames and for a range of energies,
illustrating the potential and also the challenges of using three-neutron spectroscopy to constrain the underlying interactions.
}

\FullConference{The 41st International Symposium on Lattice Field Theory (LATTICE2024)\\
 28 July - 3 August 2024\\
Liverpool, UK\\}

\begin{document}
\maketitle

\section{Introduction}
The finite-volume multi-particle formalism allows one to extract physical, infinite-volume scattering amplitudes from the results of lattice simulations. The implementation of the three-neutron case is the focus of this paper. Specifically, we follow the relativistic field theory (RFT) approach~\cite{RFT1,RFT2}, recalling that there are two steps: applying the quantization condition (QC) to lattice data and solving integral equations for infinite volume scattering amplitudes. Here we consider only the first step. 

Assuming that the two-neutron scattering amplitudes are known, the three-neutron QC gives access to the three-neutron scattering amplitude which is relevant to a variety of physical systems. These include: neutron-rich nuclei, which, as neutron density increases, become increasingly sensitive to three-neutron forces; and neutron stars, where the three-neutron force plays a part in the equation of state and nuclear matter in general. Furthermore, any multiparticle process where three neutrons can go on shell will be dependent upon the scattering amplitude. 

The implementation in this paper is based on the derivation presented in \cite{SRS1} which uses the RFT approach to derive the three-neutron QC. The body of this paper is broken into two sections, the first of which addresses some of the details particular to the case of three neutrons and how the building blocks of the QC are implemented, while the second contains results of the QC for two different relativistic frames. We conclude with a discussion of future work. %and next steps.

\section{Three-Neutron Quantization Condition}

We begin by recalling the two-particle QC. This can be written as 

\begin{equation}
    \det\left[1+K_2 F\right]=0
    \label{2pQC}
\end{equation}
where $K_2$ is the two-particle K-matrix and $F$ is a known kinematical matrix depending on the box size, $L^3$. Solutions to Eq.(\ref{2pQC}) give the finite-volume energy spectrum of the two particles. In practice, one fits these solutions to those obtained from lattice simulations to find the K-matrix and then determines the scattering amplitude from the K-matrix.

For three neutrons, there are additional complications that arise both from going to three particles and from including spin in the formalism. The QC can be written in a similar form:

\begin{equation}
    \det\left[1+K_{\rm{df},3} F_3\right]=0
    \label{3pQC}
\end{equation}
where $K_{\rm{df},3}$ is the divergence-free three-particle K-matrix, and 
\begin{equation}
    F_3=\frac{F}{3}+F\frac{1}{1-M_{2,L}G}M_{2,L}F, \qquad M_{2,L}=\frac{1}{K_2^{-1}-F}.
\end{equation}

\begin{figure}
    \centering
    \includegraphics[width=0.8\linewidth]{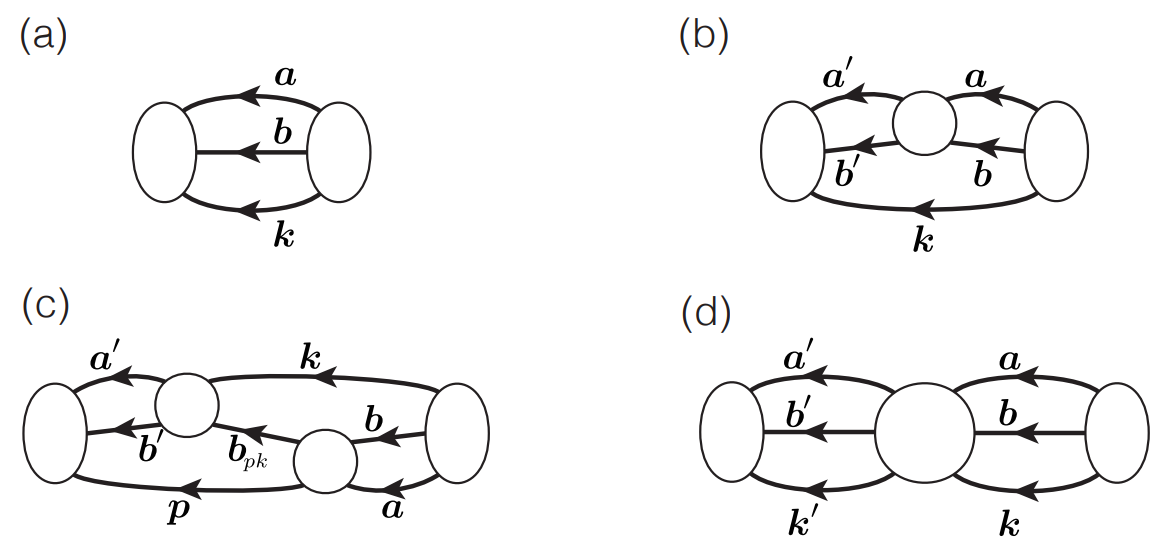}
    \caption{Components of the diagrammatic expansion of the finite volume correlator: (a) shows the three particles propagating between endcaps with an explicit insertion of the projector, $F$; (b) shows an insertion of $K_2$; (c) shows an insertion of $G$; and (d) shows an insertion of $K_{\rm{df},3}$.}
    \label{fig:dm}
\end{figure}

There are now four matrices that enter into the QC: $F$ and $K_2$ are the three-particle equivalents of the corresponding matrices from the two-particle QC, $K_{\rm{df},3}$ is defined above and $G$ is a "switch" matrix that keeps track of which two particles are involved in a given two-particle interaction. Fig.\ref{fig:dm} shows the role each of these matrices plays in the diagrammatic expansion of the correlator. 

For the results presented here, we take the limit  $K_{\rm{df},3}\xrightarrow[]{}0$. In this limit, the QC becomes $ \det\left[F_3^{-1}\right]=0$.

To implement the QC, we first need to understand the vector space on which these matrices act. For a given total energy and momentum of the three particles we have the following degrees of freedom that index the matrices: spectator momentum ($\mathbf{k}$), total ($l$) and azimuthal ($m$) orbital angular momentum of the dimer (interacting) pair, and the spin of each neutron, $\mathbf{m}_s=(m_k,m_a,m_b)$. The spectator momentum is limited by the left-hand cut (LHC) in the two-particle interaction due to single-pion exchange. For a given total momentum ($\mathbf{P}$), there is a maximum value of $\mathbf{k}$ and so the set of spectator momentum is finite. The angular momentum series is in principal infinite and must be truncated. In this work, we take $l_{max}=1$. All together, these indices lead to 32-dimensional matrices for each spectator momentum. Taking into account the antisymmetry of the three-neutron wavefunction enforces the constraint (in the case of $l_{max}=1$) that $l=s$, where $s$ is the total dimer spin, our matrices are actually 20-dimensional for each $\mathbf{k}$.

Since the number of contributing spectator momenta can be large, it is computationally advantageous to further reduce the dimensionality of the QC matrices. This can be done by projecting onto irrecucible representations (irreps) of the symmetry group in which the matrices live. For instance, if $\mathbf{P}=\frac{2 \pi}{L}(0,0,1)$, the cubic symmetry group of the lattice is broken down to the little group $C_{4v}$ and the QC matrices can be block-diagonalized in irreps. The solutions the the QC will live in definite irreps of, in this case, the double of $C_{4v}$. We can choose an irrep of interest, project the QC onto that irrep and then solve the QC to find the spectrum in that irrep. One simplification in the three-neutron case is that, since the wavefunction is fermionic, solutions of the QC must live in fermionic irreps, limiting the number of projections one must do in each little group.

The final consideration when implementing the QC is how to parameterize $K_2$. In general, $K_2$ can be expressed in terms of scattering phase shifts in the basis of total dimer angular momentum, $j$. This is true provided the different angular momentum channels do not mix. In our case the channels are: $\left[j=s=l=0\right]$, $\left[j=0,s=l=1\right]$,$\left[j=s=l=1\right]$, and $\left[j=2,s=l=1\right]$, none of which mix due to total angular momentum/parity constraints. It should be noted that $j=2$ receives a contribution from $\left[l=3,s=1\right]$ but, since we are setting $l_{max}=1$, this is assumed to be zero. 

\begin{figure}
    \centering
    \includegraphics[width=0.8\linewidth]{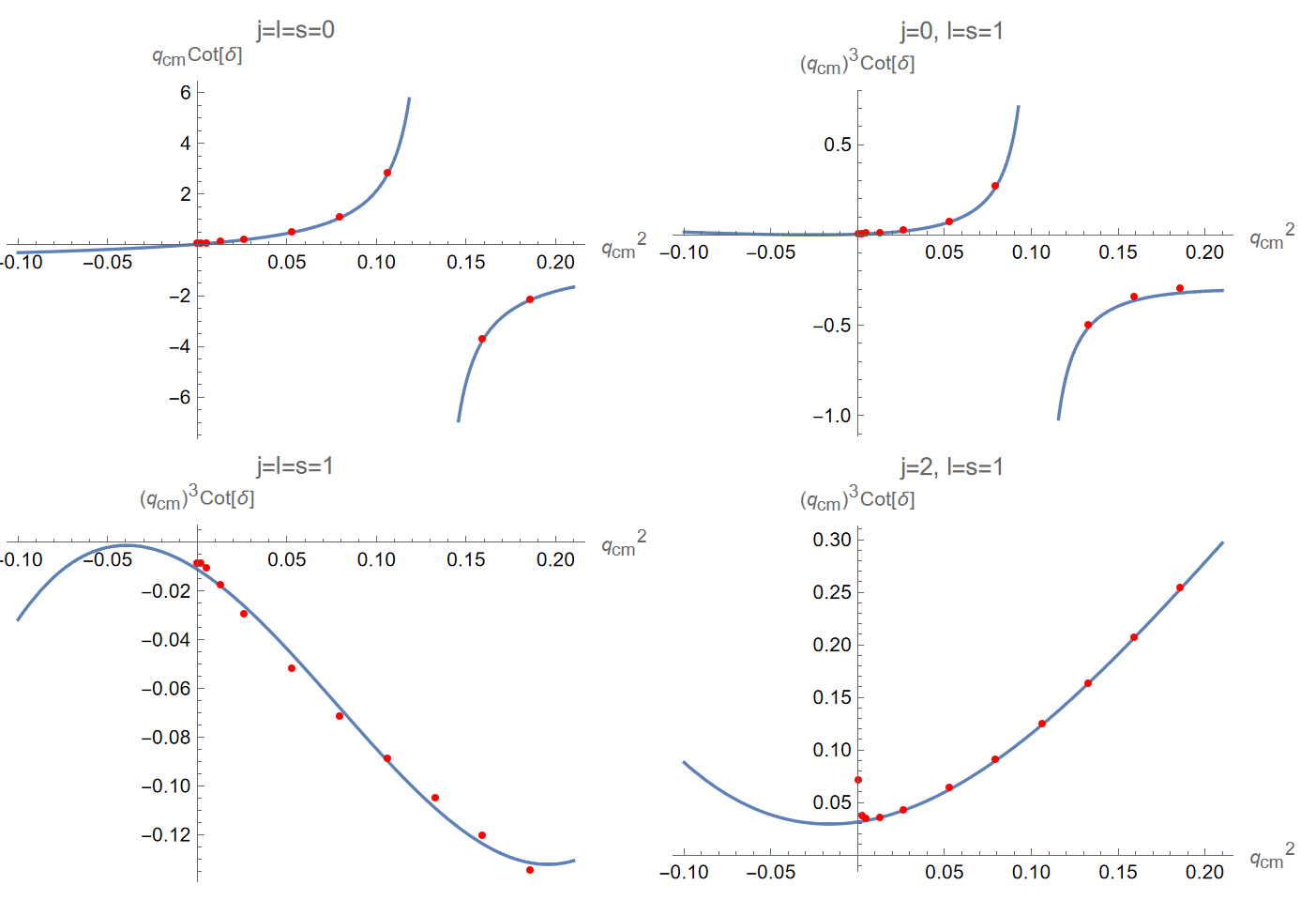}
    \caption{The functional form of the scattering phase shifts for each of the four channels in $K_2$. Data points are from Ref.~\cite{K2d}; the blue lines are our chosen forms. Quantities are expressed in units of the neutron mass.
    }
    \label{fig:sl}
\end{figure}

One can parameterize the phase shifts in different ways, such as an effective range expansion or a constant scattering length. Here we use the phase shifts shown in Fig.~\ref{fig:sl}, which are based on fits to proton-proton scattering data from Table IV of  Ref.~\cite{K2d}, where we have averaged over the models presented in that work. Our assumption is that isospin breaking is small, so that the $nn$ and $pp$ phase shifts are close. Naive fits to these models result in unobserved bound states, or other singularities in $M_2$, for $q_{cm}^2<0$. If these lie close to threshold, they can give rise to unphysical solutions to the QC. The subthreshold form of the functions in Fig.~\ref{fig:sl} have been adjusted to push such singularities away from the region of $q_{cm}^2$ that enters the QC.\footnote{%
The results presented at the lattice conference used a slightly different choice of phase shifts that did lead to singularities in $M_2$ close to threshold, which in turn lead to unphysical solutions of the QC.
These were removed by an ad hoc procedure of adding a step function to $K_2^{-1}$ such that $K_2$ was forced to be small for $q_{cm}^2 <0$. The approach presented here is preferable, as $K_2$ remains smooth, so that our choices for the phase shifts have no unphysical features in the kinematic regime of interest.
The effect of these changes on the spectrum is very small and, though included in the figures, is barely visible.}

Once we have the building blocks of the QC, the implementation steps are as follows: choose a valid energy range in which to solve the QC (must be in an interval in which only three-neutron states can go on shell),
pick a total momentum frame and irrep to project onto, solve the QC at points along the energy range and find where the determinant (or the smallest eigenvalue) crosses zero. This yields the interacting, three-neutron spectrum. In the next section we show results for the spectrum in the $\mathbf{d_P}=(0,0,1)$ frame and the $\mathbf{d_P}=(0,0,0)$ frame, where $\mathbf{P}=(2 \pi/L)\mathbf{d_P}$.

\section{Results}

We present results using the (close to physical) choice $m_\pi=0.15 m_N$, where $m_N$ is the neutron mass.
All energies below are given in neutron-mass units, such that $m_N=1$.
We choose a relatively small lattice volume, with $m_\pi L = 3$ (and thus $m_N L =20$), in order to enhance the finite-volume shifts. For physical pion masses, this corresponds to $L =4.3 \;$fm.

\subsection{$\mathbf{d_P}=(0,0,1)$}

\begin{figure}
    \centering
    \includegraphics[width=0.8\linewidth]{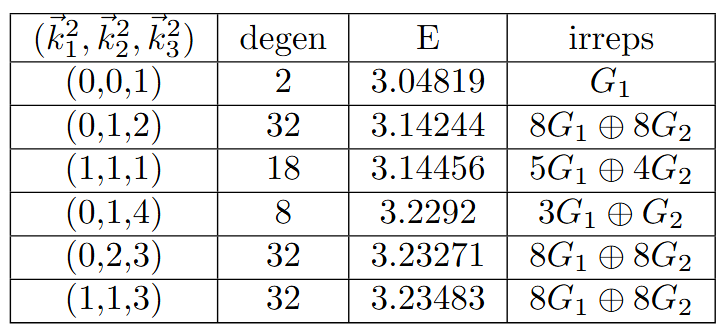}
    \caption{Lowest energy three-neutron free levels in the frame with $\mathbf{d_P}=(0,0,1)$. The first column lists the squared magnitudes of the momenta of the individual neutrons, in units of $(2\pi/L)^2$. 
    %Energies are given in units of $M_N$. 
    }
    \label{fig:p001}
\end{figure}

To solve the QC in the frame where the total momentum is proportional to $(0,0,1)$, we follow the steps described above. The doubled little group has 7 irreps, two of which are fermionic: $G_1$ and $G_2$, both of which are two-dimensional.
To identify an energy range of interest, we look at the free levels and choose those we want to track. Table~\ref{fig:p001} shows the first few free levels in this frame. We  focus on the first three levels, which have degeneracies 2, 32 and 18 respectively. Note that the second and third free levels are degenerate in the nonrelativistic (NR) limit.
  The last column in the table shows the irrep contribution for each free level. To solve the QC, we first project onto $G_1$ and scan energies from below 3.04819 to above 3.14456, finding the solutions to the QC. We then repeat the process for $G_2$.

\begin{figure}
    \centering
    \includegraphics[width=0.8\linewidth]{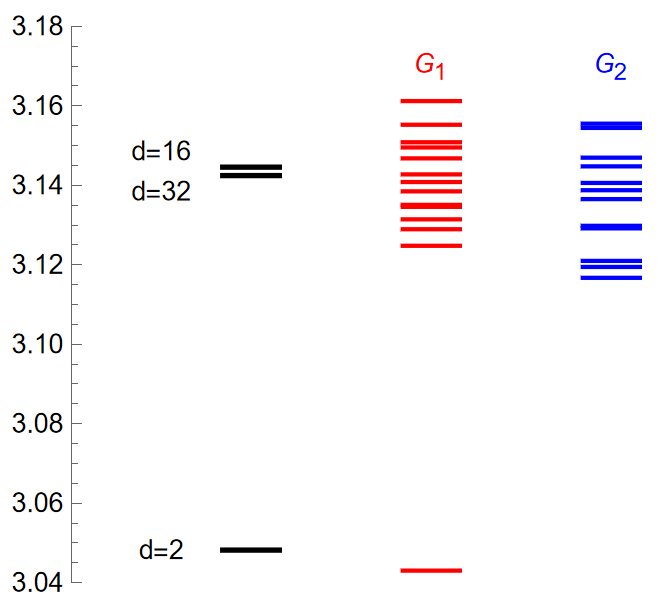}
    \caption{The energy spectrum in the $\mathbf{d_P}=(0,0,1)$ frame. The black bars are the free levels (with their degeneracies); the red and blue bars are the interacting levels separated into fermionic irreps. %, $G_1$ and $G_2$.
    }
    \label{fig:fig-p001l}
\end{figure}

Fig.~\ref{fig:fig-p001l} shows the results for the interacting spectrum. A key consistency check is that we find 14 levels in $G_1$ and 12 in $G_2$, %as we expect from the free levels. 
the same counting as for free levels. 
We do not expect additional levels due to the absence of bound states or resonances in the $nn$ system.
We observe that the interactions significantly spread the levels, although those associated with upper and lower  NR ``blocks'' do not overlap. 
It is notable that no degeneracies remain, so that the spectrum provides the maximum information possible on the underlying interactions. Nevertheless, given the densely-packed spectrum, determining these levels in a numerical simulation will be challenging.

% Although the levels do spread out in the presence of the interaction, they are still fairly tightly spaced. This may make it difficult for them to be resolved in lattice simulations. Conversely, the fact that nonrelativistically degenerate sets of levels, such as the higher two in Fig. \ref{fig:fig-p001l}, are well resolved from other sets should allow the lattice to set some bounds on the interactions.

\subsection{$\mathbf{d_P}=(0,0,0)$}
\begin{figure}
    \centering
    \includegraphics[width=0.8\linewidth]{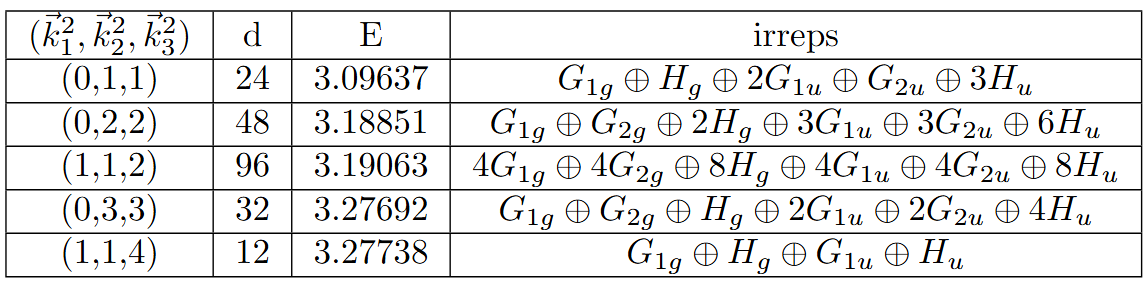}
    \caption{Lowest energy three-neutron free levels in the frame with $\mathbf{d_P}=(0,0,0)$.
    Notation as in Table~\ref{fig:p001}.}
    \label{fig:p000}
\end{figure}

We follow the same steps as above for the $\mathbf{d_P}=(0,0,0)$ case. Now, however, the symmetry group is the full doubled octahedral group with six fermionic irreps. The ``$G$'' and ``$H$" irreps have, respectively, dimensions 2 and 4. The first few free levels are listed in Table~\ref{fig:p000} and one can see that the degeneracies of the free levels are larger and have contributions from many more irreps than for $\mathbf{d_P}=(0,0,1)$.

\begin{figure}
    \centering
    \includegraphics[width=0.8\linewidth]{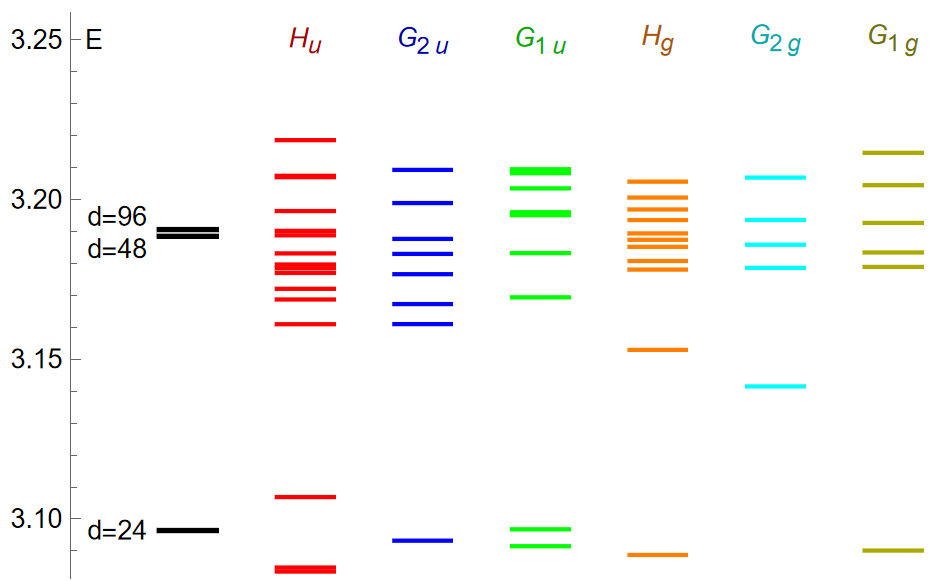}
    \caption{The energy spectrum in the $\mathbf{d_P}=(0,0,0)$ frame. The black bars are the free levels (with their degeneracies); the colored bars are the interacting levels separated into fermionic irreps. % $H_u$, $G_{2u}$, $G_{1u}$, $H_g$, $G_{2g}$ and $G_{1g}$.
    }
    \label{fig:p000l}
\end{figure}

Results for the interacting spectrum across the first three sets of degenerate free levels are shown in Fig.~\ref{fig:p000l}.. The results clearly illustrate the advantage of decomposing levels into irreps. As in the $(0,0,1)$ case, most of the irreps have tightly-spaced interacting levels that could be hard to resolve on the lattice. $G_{1g}$ and $G_{2g}$, however, are comparatively well-resolved and could provide an access point for the lattice in determining the interactions.

\section{Conclusion and Future Work}

In the preceding sections, we have shown how the RFT QC formalism can be used to find the three-neutron spectrum, including at this stage only  two-particle interactions. The results show that for a reasonable parameterization of the two-neutron K-matrix, free levels spread out but still remain fairly tightly spaced in most irreps. The next step is to implement the QC with the three-neutron K-matrix, $K_{\rm{df},3}$. A preliminary analysis indicates that, to leading order, $K_{\rm{df},3}$ only leads to shifts in a subset of levels in the spectrum, which may make it more feasible to isolate three-particle effects. Ultimately, we await the lattice results to which we can apply this machinery.

\section*{Acknowledgements}
We thank Max Hansen, Fernando Romero-L\'opez, and Akaki Rusetsky for useful comments and discussions.
This work is supported in part by the U.S. Department of Energy grant No.~DE-SC0011637. 
This work contributes to the goals of the USDOE ExoHad Topical Collaboration, contract DE-SC0023598.


\begin{thebibliography}{99}
\bibitem{RFT1}
M.~T.~Hansen and S.~R.~Sharpe,
{\it ``Relativistic, model-independent, three-particle quantization condition,''}
Phys. Rev. D \textbf{90}, no.11, 116003 (2014)
%doi:10.1103/PhysRevD.90.116003
[arXiv:1408.5933 [hep-lat]].
%242 citations counted in INSPIRE as of 08 Oct 2024
\bibitem{RFT2}
M.~T.~Hansen and S.~R.~Sharpe,
{\it ``Expressing the three-particle finite-volume spectrum in terms of the three-to-three scattering amplitude,''}
Phys. Rev. D \textbf{92}, no.11, 114509 (2015)
%doi:10.1103/PhysRevD.92.114509
[arXiv:1504.04248 [hep-lat]].
%232 citations counted in INSPIRE as of 08 Oct 2024
\bibitem{SRS1}
Z.~T.~Draper, M.~T.~Hansen, F.~Romero-L\'opez and S.~R.~Sharpe,
{\it ``Three relativistic neutrons in a finite volume,''}
JHEP \textbf{07}, 226 (2023)
%doi:10.1007/JHEP07(2023)226
[arXiv:2303.10219 [hep-lat]].
%20 citations counted in INSPIRE as of 08 Oct 2024
%Draper, Z.T., Hansen, M.T., Romero-López, F., and Sharpe, S.R. \it Three relativistic neutrons in a finite volume. \rm J. High Energ. Phys. 2023, 226 (2023). https://doi.org/10.1007/JHEP07(2023)226
\bibitem{K2d}
V.~G.~J.~Stoks, R.~A.~M.~Klomp, C.~P.~F.~Terheggen and J.~J.~de Swart,
{\it ``Construction of high quality N N potential models,''}
Phys. Rev. C \textbf{49}, 2950-2962 (1994)
%doi:10.1103/PhysRevC.49.2950
[arXiv:nucl-th/9406039 [nucl-th]].
%1451 citations counted in INSPIRE as of 08 Oct 2024
%Stoks, V.G.J., Klomp, R.A.M., Terheggen, C. P. F., and  de Swart, J. J. \it Construction of high-quality NN potential models. \rm Phys. Rev. C 49, 2950 (1994)
\end{thebibliography}
\end{document}